\documentclass[a4paper]{jpconf}
\usepackage{graphicx}
\begin{document}

\newcommand*\aap{A\&A}
\let\astap=\aap
\newcommand*\aapr{A\&A~Rev.}
\newcommand*\aaps{A\&AS}
\newcommand*\actaa{Acta Astron.}
\newcommand*\aj{AJ}
\newcommand*\ao{Appl.~Opt.}
\let\applopt\ao
\newcommand*\apj{ApJ}
\newcommand*\apjl{ApJ}
\let\apjlett\apjl
\newcommand*\apjs{ApJS}
\let\apjsupp\apjs
\newcommand*\aplett{Astrophys.~Lett.}
\newcommand*\apspr{Astrophys.~Space~Phys.~Res.}
\newcommand*\apss{Ap\&SS}
\newcommand*\araa{ARA\&A}
\newcommand*\azh{AZh}
\newcommand*\baas{BAAS}
\newcommand*\bac{Bull. astr. Inst. Czechosl.}
\newcommand*\bain{Bull.~Astron.~Inst.~Netherlands}
\newcommand*\caa{Chinese Astron. Astrophys.}
\newcommand*\cjaa{Chinese J. Astron. Astrophys.}
\newcommand*\fcp{Fund.~Cosmic~Phys.}
\newcommand*\gca{Geochim.~Cosmochim.~Acta}
\newcommand*\grl{Geophys.~Res.~Lett.}
\newcommand*\iaucirc{IAU~Circ.}
\newcommand*\icarus{Icarus}
\newcommand*\jcap{J. Cosmology Astropart. Phys.}
\newcommand*\jcp{J.~Chem.~Phys.}
\newcommand*\jgr{J.~Geophys.~Res.}
\newcommand*\jqsrt{J.~Quant.~Spectr.~Rad.~Transf.}
\newcommand*\jrasc{JRASC}
\newcommand*\memras{MmRAS}
\newcommand*\memsai{Mem.~Soc.~Astron.~Italiana}
\newcommand*\mnras{MNRAS}
\newcommand*\na{New A}
\newcommand*\nar{New A Rev.}
\newcommand*\nat{Nature}
\newcommand*\nphysa{Nucl.~Phys.~A}
\newcommand*\pasa{PASA}
\newcommand*\pasj{PASJ}
\newcommand*\pasp{PASP}
\newcommand*\physrep{Phys.~Rep.}
\newcommand*\physscr{Phys.~Scr}
\newcommand*\planss{Planet.~Space~Sci.}
\newcommand*\pra{Phys.~Rev.~A}
\newcommand*\prb{Phys.~Rev.~B}
\newcommand*\prc{Phys.~Rev.~C}
\newcommand*\prd{Phys.~Rev.~D}
\newcommand*\pre{Phys.~Rev.~E}
\newcommand*\prl{Phys.~Rev.~Lett.}
\newcommand*\procspie{Proc.~SPIE}
\newcommand*\qjras{QJRAS}
\newcommand*\rmxaa{Rev. Mexicana Astron. Astrofis.}
\newcommand*\skytel{S\&T}
\newcommand*\solphys{Sol.~Phys.}
\newcommand*\sovast{Soviet~Ast.}
\newcommand*\ssr{Space~Sci.~Rev.}
\newcommand*\zap{ZAp}

\title{Nucleosynthesis in the first massive stars}

\author{Arthur Choplin$^1$, Georges Meynet$^1$, Andr\'e Maeder$^1$, Raphael Hirschi$^2$, Cristina Chiappini$^3$}

\address{$^1$Geneva Observatory, University of Geneva, Maillettes 51, CH-1290 Sauverny, Switzerland}

\address{$^2$Astrophysics group, Lennard-Jones Laboratories, EPSAM, Keele University, ST5 5BG, Staffordshire, UK}

\address{$^3$Leibniz-Institut fur Astrophysik Potsdam (AIP), An der Sternwarte 16, 14482 Potsdam, Germany}

\ead{georges.meynet@unige.ch}

\begin{abstract}
The nucleosynthesis in the first massive stars may be constrained by observing the surface composition of long-lived very iron-poor stars born around 10 billion years ago from material enriched by their ejecta.
Many interesting clues on physical processes having occurred in the first stars  can be obtained just based on nuclear aspects. Two facts are particularly clear, 1) in these first massive stars, mixing must have occurred 
between the H-burning and the He-burning zone during their nuclear lifetimes; 2) only the outer layers of these massive stars have enriched the material from which the very iron-poor stars,
observed today in the halo, have formed. These two basic requirements can be obtained by rotating stellar models at very low metallicity. In the present paper, we discuss the arguments supporting this view and
illustrates the sensitivity of the results concerning the [Mg/Al] ratio on the rate of the reaction $^{23}$Na(p,$\gamma$)$^{24}$Mg.
\end{abstract}

\section{Some puzzling abundance patterns}

If a particularly imaginative theoretician would have imagined the composition of very metal-poor stars formed from the ejecta of the first stellar generations, he would likely never have succeeded to
predict the very strange abundance patterns shown by these stars. The fact that they present very low iron is per se not a surprise, but the fact that they present very strong overabundances of carbon with respect
to iron is at first sight astonishing. Moreover, other light elements like nitrogen, oxygen, sodium, magnesium and aluminum also show excesses with respect to iron\footnote{Actually these are excesses compared to the abundances of these elements
with respect to iron in the Sun.}. So the question is which kind of sources can predict such abundance patterns? This is the main topic of the present discussion. However before to address that specific question, it is worthwhile to be aware
of a few facts. 

The stars we are speaking about are named the Carbon-Enhanced Metal Poor stars. These stars have clearly much larger [C/Fe] than the bulk of the normal halo stars (Aoki et al. 2007). Some of them present significant excesses in r- and s-process elements. In that case they are classified as CEMP-r, CEMP-s or CEMP-r/s stars. Some of them present no or only very small excesses in neutron capture elements and are named CEMP-no stars. Presently 46 stars of this type are known (see table 2 in Maeder \& Meynet 2015). The sample increases slowly especially at the very low metallicity end. A way to illustrate this is to note that the number of stars mentioned in the review by Beers \& Christlieb (2005) with a metallicity [Fe/H] below -4 was 6. Ten years later, in 2015, only two stars have been added in that metallicity range, which rises the number to 8.
The 8 stars  are all CEMP-no stars except one. 

How did these 7 very metal poor stars form? Why are they carbon-rich? Could these abundances be built by the star itself? If not, 
what was the nature of the nucleosynthetic sources that build up their abundance pattern?  These questions are the main topics of the next section.

\section{Some clues from nuclear considerations}

\begin{figure}
\begin{center}
\includegraphics[width=18.5pc]{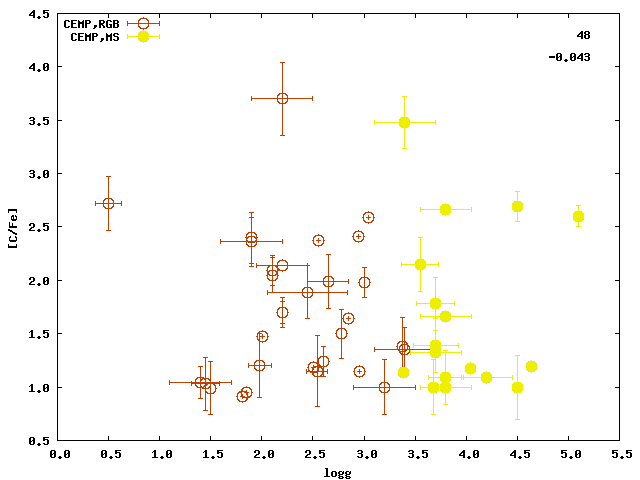}\includegraphics[width=18.5pc]{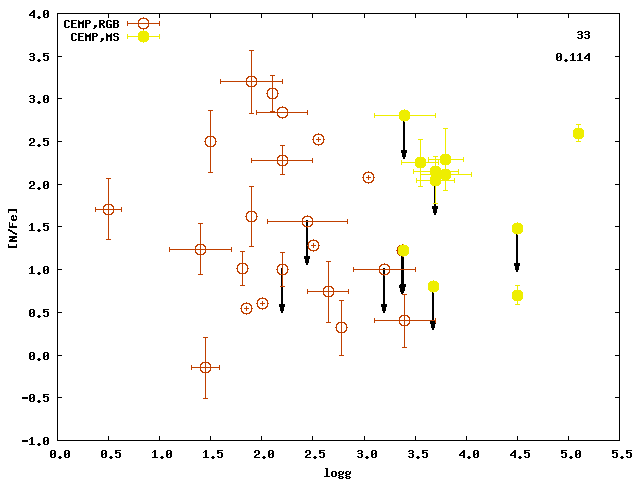}
\end{center}
\caption{\label{MSrg}{\it Left panel:} CEMP stars with [Fe/H] $<$ -2.0, [C/Fe] $>$ 0.9 and [Ba/Fe] $<$ 0.9 plotted in the plane [C/Fe] versus the surface gravity of the stars. Different colors are for Main-Sequence and red giant stars. {\it Right panel:} Same stars as those plotted in the left panel that have measured values for the [N/Fe] ratio. The database SAGA has been used to make this plot (see Suda et al. 2008).}
\end{figure}

CEMP-no stars are found among Main-Sequence and evolved red giant stars. In the left panel of Fig.~\ref{MSrg}, we can see that the scatter in carbon overabundance is more or less the same for non-evolved Main-Sequence (MS) stars and evolved red giant (RG) stars. This supports the idea that the carbon enrichment is not just an enrichment of only the outer layers of the stars acquired by the accretion of a thin enriched layer. Would it be the case, then how to explain this similarity between MS and RG stars? At the red giant stage, a deep convective envelope produces a rapid mixing and any superficial enrichment would be rapidly diluted and made either unobservable or at least much smaller than during the MS phase. So what we see at the surface has great chance to be the bulk composition of the whole star, not just of its surface.

Stars plotted in Fig.~\ref{MSrg} are very small mass stars (around 0.6-0.8 M$_\odot$). These stars cannot produce any carbon in the stellar evolutionary phases in which they are observed. Thus they must have inherited these abundances from the protostellar cloud from which they formed.
The source stars, that produced the enrichment of this protostellar cloud are very likely massive stars. What support this statement is the very low iron content in CEMP-no stars. Actually the iron content is so low that the pocket must have been enriched by a unique or may be two events able to enrich their surrounding in short enough timescales for avoiding the pocket to be enriched by many other massive stars that would otherwise increase the iron content rapidly above the observed level.

The CEMP-no stars clearly show the signature of H- and He-burning. 
Concerning H-burning, one (among many others) very clear signature comes from the fact than most CEMP-no stars present nitrogen overabundances.
This is shown in the right panel of Fig.~\ref{MSrg}. 
As for the [C/Fe] ratio, there is no very relevant differences between MS and RG stars. During the RG phase, some nitrogen surface enrichment may occur  due to the concomitant effects of the CNO cycle operating in the H-burning shell and the dredge-up due to convection.

There are also very clear signatures of the operation of He-burning reactions. The presence, at the surface of CEMP-no stars, of high abundances of carbon, element produced by the triple $\alpha$ reaction, is one of them (see the left panel of Fig.~\ref{MSrg}).
Are CEMP-no stars made of a mixture of H- and He-burning processed material? Actually, while such a mixture is indeed required, it is still not sufficient. To see that, let us make a very simple thought experiment. Let us consider for instance some material having initially a very low metal content, let us say to fix the idea having a mass fraction of heavy elements (Z) equals to 10$^{-4}$. If we assume that the heavy elements are distributed like in the Sun, then the initial abundance of CNO elements in mass fraction would correspond to about 3/4 of Z, thus of the order of 10$^{-4}$. Since the CNO cycle mainly transforms the carbon and oxygen into nitrogen, at the end of the process, the nitrogen abundance would also be around 10$^{-4}$. The mass fraction of iron would not changed and will stay around 1/20 of the initial Z, that means 5 10$^{-6}$, so that in CNO-processed material one expects a [N/Fe] at most around 1.3 - 0.3 = 1 dex (since the brackets mean relatively to the sun and (N/Fe)$_{\odot} \simeq $ 0.3 dex). Actually this is an upper limit since we assumed a complete transformation of carbon and oxygen and we did not account for any dilution effect when this material is mixed with the material processed by He-burning.

Now let us consider the star G77-61 for instance which has a surface gravity equal to 5.1 (Beers et al. 2007) and a [Fe/H] equals to -4. That star is the MS star shown at the right in both panels of Fig.~\ref{MSrg}. 
Let us assume that this star was made from the ejecta of a source star having also a [Fe/H] equal to -4 and a solar distribution of the heavy elements. As explained above, one would then obtain an upper limit for the [N/Fe] equals to 1, but this is much too small with respect to the observed value of 2.6. It means that the above scenario is unable to account for the observed N overabundance. Something is missing. 
We could argue that the source star began with a non solar initial composition, for instance which much larger carbon initial abundance but in that case, it would just shift the problem: how to explain the high carbon abundance in that previous generation of stars? This simple example shows that just mixing together material processed by H- and He-burning is not sufficient. Actually, in order to produce the large amounts of nitrogen, it is necessary to assume that some mixing operates between these two nuclear burning regions. If some carbon produced in the He-core diffuses into the H-burning region then this would produce much larger amounts of nitrogen. 
Thus we can conclude that the abundance patterns of CEMP-no stars provide some clues for the existence of a mixing mechanism in the source stars having occurred during their nuclear lifetimes.

There is still one more conclusion that we can deduce just from nuclear physics arguments. We just saw that the abundance patterns of CEMP-no stars present the signature of He-burning, but this signature
does not overwhelm the one of the H-burning process.  Too much He-burning material would produce very high [C/N] ratios
due to the fact that carbon is so much more abundant in the He-burning region than in the H-burning one.
This would make impossible to understand a [C/N] ratio of about 0 in the case G77-61. It would also be in complete contradiction with the very low $^{12}$C/$^{13}$C ratio measured at the surface of this star which is around 5.
This value is equal to the value expected from CNO processing alone (with some carbon produced by He-burning injected into the H-burning shell to account for the high carbon abundances). 
If too much material from the He-burning region would contribute in making the material from which this star formed, then one would obtain high [C/N] and $^{12}$C/$^{13}$C ratios that are not observed. 
This supports the view that no or only a very small amount of the He-burning zone has enriched that material. 

Let us briefly make the synthesis of what we have learnt at that point: nuclear physics arguments tells us that CEMP-no stars are made of material having been processed by H- and He-burning nuclear reactions with
some exchanges between them and that only small amounts of He-burning regions has been injected into this material.

\section{From nuclear considerations to a stellar model}

Let us now see how the above conditions can be realized in a star. Having material processed by H- and He-burning nuclear reactions is per se not constraining since these two nuclear networks
are active in any massive stars. Now we need mixing to occur between these two zones. Meynet \& Maeder (2002a), Hirschi (2007) have proposed that rotation might trigger this mixing, producing large amounts of nitrogen at low metallicity. Chiappini (2013) in a recent review discusses many other indications supporting rotational mixing from nucleosynthetic arguments. In brief, the arguments favoring rotational mixing, and a more efficient mixing at low metallicity are the following: 
first the fraction of rapidly rotating stars is larger when the metallicity is decreasing (Maeder et al. 1999, Martayan  et al. 2007). Second the rotational mixing is more efficient, all other parameters being kept the same, when the metallicity is lower (Maeder \& Meynet 2001). This is due to the fact that stars at low metallicity are more compact.  Also,  in metal-poor massive stars, the distance between the H- and He-burning zone decreases since the lack of CNO elements implies that the H-burning shell occurs at higher temperature and thus in a zone nearer from the He-burning core. All these considerations produce the interesting situation that the mixing which is indeed needed at very low metallicity to explain the CEMP-no stars, occurs only at low or very low metallicity. It is absent at solar metallicity for instance. This is quite consistent with the fact that at solar metallicity there are no evidences for processes requiring such mixing like the production of primary nitrogen\footnote{Primary nitrogen is produced from carbon and oxygen synthesized by the star in its He-burning zone, while secondary nitrogen is produced from carbon and oxygen already present in the star at its formation and thus resulting from one (or a few) previous generation(s) of star.}.

Another interesting aspect of this kind of models is that they might also produce conditions for triggering stellar winds even if the source massive star had a very low initial metallicity. This comes from the fact that the rotational mixing can enrich the surface in CNO elements that might enhance the opacity of the outer layers during the red supergiant phase and thus triggers important mass losses through radiatively driven stellar winds (Meynet \& Maeder 2002b). The fact that the matter is ejected with a low velocity may help star formation to occur rapidly from a mixture of such stellar winds and ISM. The matter ejected at high velocity during the supernova ejecta might on the other hand escape from the potential well. So we see that rotation might produce both interesting abundance patterns and at the same time favorable conditions for star formation to occur rapidly from a mixture of the ejected material with some ISM.

Such a model requires, on the other hand, that the metallicity of the source stars be equal or higher (in case of dilution) than the metallicity of the CEMP-no stars, because the wind is not enriched in iron.
Thus, in that case, the source star would not be a Pop III star but a very metal poor massive star of likely second generation. Another possibility would be that the source star would be primordial. In that case, it is needed that some ejection of iron occurs at the moment of the faint supernova event as proposed by Umeda \& Nomoto (2005). This last scenario could also occur in case of a non-zero metallicity stars. 

At the moment it is difficult  to disentangle between these two possibilities. One can wonder whether in case of Pop III star, there would still be possible to account for the
small amounts of Strontium and Barium that are observed at the surface of some CEMP-no stars. If these elements are produced by the r-process then this would indicate
some contribution from the supernova explosion and it could explain their presence even starting from a Pop III star.
If they are produced by the s-process then likely the source star was not a Pop III stars, since in that case the production of s-process elements
might be nearly zero due to the lack of iron seeds (it might be not completely zero due to the fact that other elements could serve as seeds for neutron captures at low metallicity). Further investigations are needed to explore this point.

\section{Nuclear reaction rate uncertainties}

\begin{figure}
\begin{center}
\includegraphics[width=30pc]{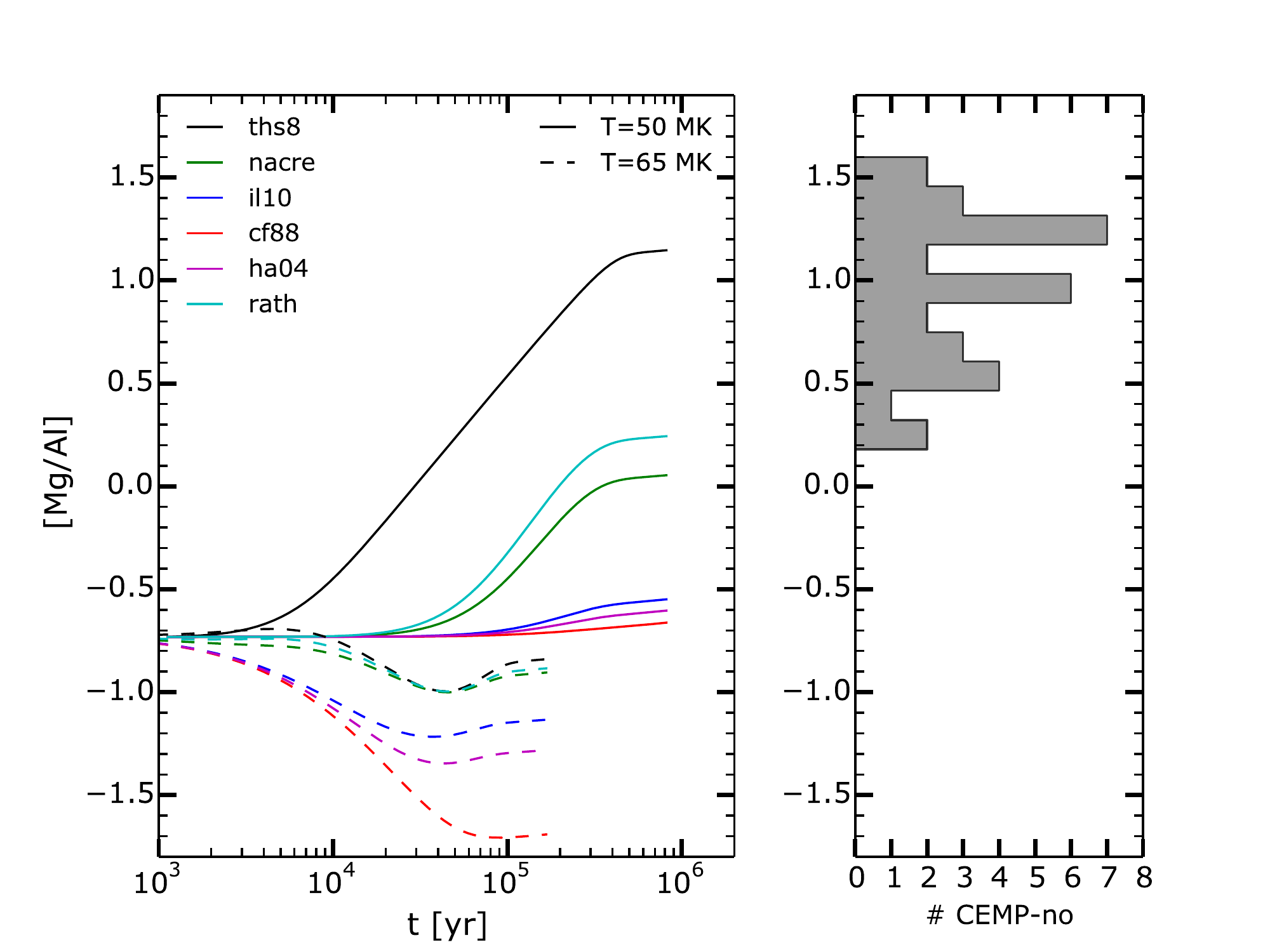}
\end{center}
\caption{\label{mgal}{\it Left panel:} [Mg/Al] ratio as a function of time in a Hydrogen burning box where we inject $^{12}$C, $^{16}$O and $^{22}$Ne.  6 rates are tested for the reaction $^{23}$Na(p,$\gamma$)$^{24}$Mg at $T_6 = 50$ and 65. Tested rates are from Cyburt et al. (2010) (ths8), Angulo et al. (1999) (nacre), Iliadis et al. (2010) (il10), Caughlan \& Fowler (1988) (cf88), Hale et al. (2004) (ha04) and Rauscher \& Thielemann (2000) (rath).   {\it Right panel:} observed [Mg/Al] distribution for CEMP-no stars.}
\end{figure}

The results of models for the source stars depends on many physical ingredients such as the way of treating the rotational mixing for instance. Among the physical ingredients, there is also the impact of some
uncertain nuclear reaction rates. In order to see more clearly what could be the impact of a given nuclear reaction rate, it is common to study the nucleosynthesis in a box at a fixed temperature and density and see
how the abundances obtained change when different nuclear reaction rates are used. In Fig.~\ref{mgal}, we show the result of such an experiment. The density is set to $1$ g cm$^{-3}$. The initial abundances in the box are taken from the Hydrogen burning shell of a $20$ $M_{\odot}$ model at the beginning of the core Helium burning phase. The computation is done at two temperatures $T_6 = 50$ (plain lines) and $T_6 = 65$ (dashed lines) which are typical temperature of H-shell burning. To model the mixing occurring between
the He- and H-burning region, $^{12}$C, $^{16}$O and $^{22}$Ne are injected into the box at constant rates $R_{^{12}C} = 10^{-8}$ yr$^{-1}$, $R_{^{16}O} = 10^{-8}$ yr$^{-1}$ and $R_{^{22}Ne} = 10^{-10}$ yr$^{-1}$. Those rates represents the mass fraction we add per year of a given element. We chose those species because they correspond in general to the most abundant species entering by mixing from the He-burning core to the H-burning shell of a low metallicity massive stellar model. 
The nitrogen produced in the H-shell may diffuse at its turn into the He-core. 
The chain $^{14}$N($\alpha,\gamma$)$^{18}$F(,$e^+ \nu_e$)$^{18}$O($\alpha,\gamma$)$^{22}$Ne allows the synthesis of a large amount of $^{22}$Ne, which also diffuses into the H-burning shell. We note that the amount of $^{22}$Ne entering in the H-burning shell is lower (at least 2 dex lower than the amount of $^{12}$C and $^{16}$O) so that the rate $R_{^{22}Ne}$ is taken 100 times smaller than the rates $R_{^{12}C}$ and $R_{^{16}O}$.
Six rates for the reaction $^{23}$Na(p,$\gamma$)$^{24}$Mg are tested, taken from the database JINA REACLIB database (Cyburt et al. 2010).

The duration of the simulation depends on the temperature : the Hydrogen is consumed more rapidly if higher temperatures are chosen. The injection of  $^{22}$Ne ignites the Ne-Na chain and also the Mg-Al chain, if the temperature is high enough. $T_6 = 50$ (plain lines) is a too low temperature for the Magnesium to burn into Aluminum. Magnesium is then only synthesized, essentially through the reaction $^{23}$Na(p,$\gamma$)$^{24}$Mg. At $T_6 = 65$, the reaction $^{26}$Mg(p,$\gamma$)$^{27}$Al becomes significant so that Magnesium is also destroyed, overall reducing the [Mg/Al] ratios compared to the $T_6 = 50$ case.

These numerical experiments can be used to indicate which rates are still not known with a sufficient  accuracy. Here we see that 
still large uncertainties pertain the rate of the reaction $^{23}$Na(p,$\gamma$)$^{24}$Mg in the low temperature range ($T_6 = 50 - 65$), producing therefore very different outcomes for the [Mg/Al] ratios. 
On the right panel of Fig.~\ref{MSrg}, the distribution of the observed [Mg/Al] ratios in CEMP-no stars is shown. We see that the observed scatter is smaller than the scatter due to the uncertainties in the nuclear physics.
Such test cannot be used to constrain the nuclear reaction rates, since the final result depends on some arbitrary chosen ingredients of the numerical experiment, but at least it clearly shows the needs from improvements on the side
of the nuclear physics before some relevant conclusions can be drawn from comparisons between models and observations. 


\nocite{angulo99,aoki07, beers05, beers07, caughlan88, Chiappini2013, 
cyburt10, hale04, hirschi07, iliadis10, MaederGrebelM, maeder01,maeder15, Martayan2007, MM2002, MMcno, rauscher00, suda08, umeda05}

\section*{References}
\bibliography{GMEYNET.bib}


\end{document}